\documentclass[11pt, oneside]{article}   	
\usepackage{amssymb}
\usepackage{tabulary}

\usepackage[table]{xcolor}
\usepackage{booktabs}
\usepackage[percent]{overpic}
\usepackage{rotating}
\usepackage{array}
\usepackage{comment}
\usepackage{colortbl}
\usepackage[USenglish]{babel}
\usepackage[useregional]{datetime2}
\usepackage{setspace}
\usepackage{pdfpages}
\usepackage{enumerate}
\usepackage[inline]{enumitem}
\usepackage{setspace}
\usepackage[pdftex]{hyperref}
\usepackage{url}
\hypersetup{ hidelinks, }


\makeatletter
\newenvironment{chapquote}[2][2em]
  {\setlength{\@tempdima}{#1}%
   \def\chapquote@author{#2}%
   \parshape 1 \@tempdima \dimexpr\textwidth-2\@tempdima\relax%
   \itshape}
  {\par\normalfont\hfill--\ \chapquote@author\hspace*{\@tempdima}\par\bigskip}
\makeatother

\newcommand{\mydate}{\DTMdisplaydate{2015}{12}{2}{-1}}

\begin{document}
\begin{titlepage}
\selectlanguage{USenglish}
\newcommand{\HRule}{\rule{\linewidth}{0.5mm}}
\center

\textsc{\LARGE University of Victoria}\\[1.5cm]
\textsc{\Large Department of Computer Science}\\[0.5cm]
\textsc{\large PhD Candidacy Proposal}\\[0.5cm]

\HRule \\[0.4cm]
{ \huge \bfseries Beyond Agile: Studying The Participatory Process in Software Development}\\[0.4cm]
\HRule \\[1.5cm]

\begin{minipage}{0.4\textwidth}
\begin{flushleft} \Large
\emph{Author:}\\
Alexey Zagalsky
\end{flushleft}
\end{minipage}
~
\begin{minipage}{0.4\textwidth}
\begin{flushright} \Large
\emph{Supervisor:} \\
Dr. Margaret-Anne Storey
\end{flushright}
\end{minipage}\\[4cm]

\Large \emph{Committee Members::}\\
Dr. Arie van Deursen\\
Dr. Leif Singer\\[3cm]

{\large \mydate}\\[3cm]
\vfill
\end{titlepage}

\section*{Preface}
\begin{chapquote}{Theodore Roosevelt}
``Nothing in the world is worth having or worth doing unless it means effort, pain, difficulty.''
\end{chapquote}

Writing this research proposal was not easy for me, in fact it was quite challenging. Beyond overcoming the regular writer's block, I think the real challenge was actually attempting to piece all the ``puzzle'' pieces together---all the work done so far---and writing in a cohesive manner. I guess the scary part was not being certain if and how the pieces would connect to each other. By doing this process, I realized that writing the research proposal benefits me more than I initially anticipated. Not only it serves as a progress report to my supervising committee, but it forced me to stop, and reflect on the steps I've taken. To digest and re-examine the work I've done, enabling me to see that there is unique and real value to my work, and see what that value is.

Does this mean I got everything figured out and planned from now on forward? No, but now I have some ground to stand on, and a direction to follow.

\newpage

\tableofcontents

\newpage
\begin{onehalfspacing}

%!TEX root = candidacy.tex

\section{Introduction} % (fold)
\label{sec:introduction}

Software development is a knowledge-building process~\cite{naur1985programming,Robillard1999}. Software is built with the \textit{tacit knowledge} in the developers' head, and the \textit{externalized knowledge} embodied in the development tools, channels, and project artifacts~\cite{naur1985programming}. However, this knowledge-building process is evolving. The formation of communities of practice and the emergence of socially enabled tools and channels has led to a paradigm shift in software development~\cite{Storey2014} and the creation of a highly tuned \textit{participatory culture}. Modern software development is collaborative, global, and large in scale (e.g., the \emph{Ruby on Rails} project has more than 2800 contributors\footnote{\url{https://github.com/rails/rails}}). Many developers now contribute to multiple projects, and as a result, project boundaries blur, not just in terms of their architecture and how they are used, but also in terms of how they are authored. Indeed, we see a participatory culture~\cite{jenkins2009confronting} forming within many development communities of practice~\cite{Wenger2000} where developers want to engage with, learn from and co-create with other developers. Developers care not just about the code they need to write, but also about the skills they acquire~\cite{singer2013mutual}, the contributions they make and the connections they establish with other developers. These activities, in turn, lead to more collaborative software development opportunities.

We see that the collaborative and participatory nature of software development continues to evolve, shape and be shaped by communication channels that are used by developer communities of practice~\cite{Lanubile2013}---both by traditional communication channels (e.g., telephone, in-person interactions), as well as social features that may be standalone or integrated with other development tools (e.g., email, chat, and forums). Within a community of practice, software is a combination of the externalized knowledge (e.g., code, documentation, history of activities) as well as the tacit knowledge that resides in the community members' heads (e.g., experience of when to use an API, or design constraints that are not written down). Communication channels and development tools support developers in forming and sharing both externalized and tacit knowledge in a highly collaborative manner.

\begin{figure}[h!]
    \begin{center}

        \includegraphics[width=\textwidth]{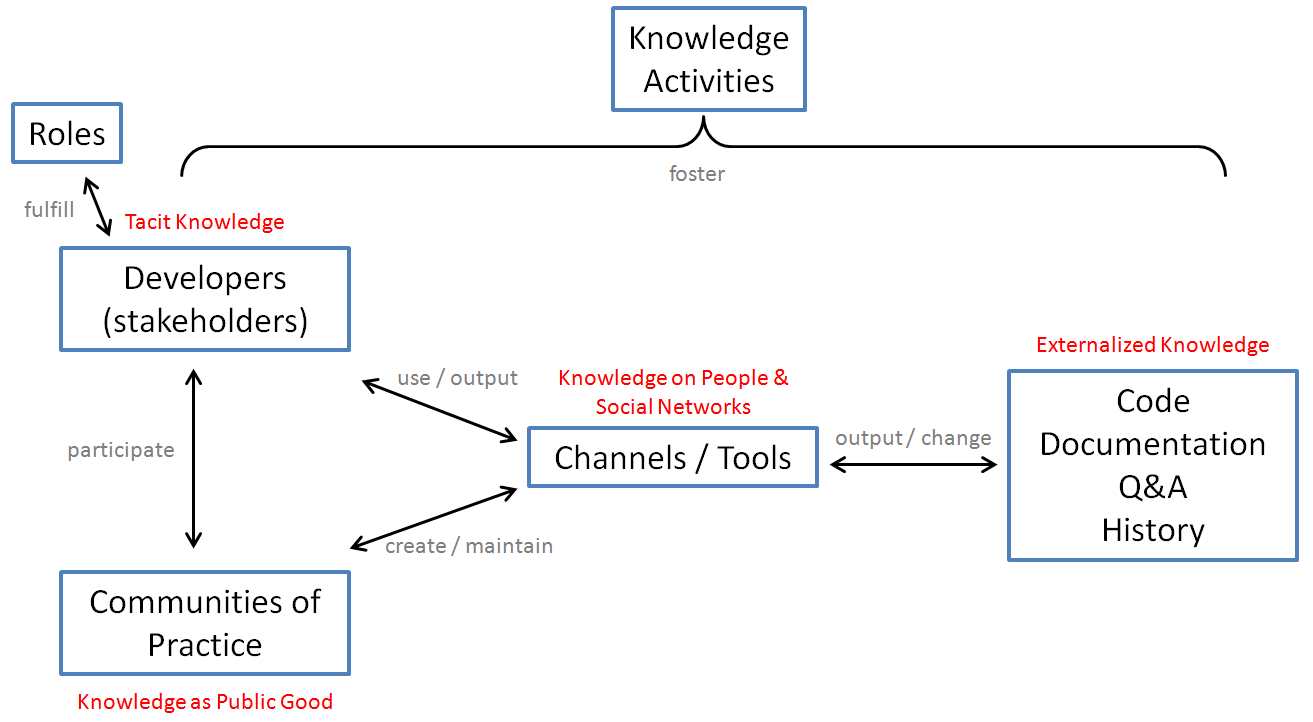}
        \caption{Knowledge theory in software development.}
        \label{fig:knowledge_diagram}
    \end{center}

\end{figure}

However, not much is known about the impact this participatory culture is having on software development practices nor on the knowledge-building processes. The research community has been studying the tools and communication channels used by developers~\cite{Singer2014,Storey2014} in an effort to broaden our understanding. But, studying only the tools and channels can only provide a narrow perspective. We believe that software development has evolved beyond the agile process, into a \textit{Participatory Process}---A knowledge building process which is characterized by the (1) \textbf{knowledge activities} and \textbf{actions}, (2) \textbf{stakeholder roles}, and (3) is enabled by socially enhanced \textbf{tools and communication channels} (as illustrated in Fig.~\ref{fig:knowledge_diagram}). Thus, it is important to gain an understanding of each one of the components, and the way these components interact and shape each other. Beyond studying how software engineering is practiced, it is equally important to consider the software engineering \textbf{research methods and practices}, and software engineering \textbf{education paradigm}, as they feed and shape each other, and are part of the same ecosystem.

\subsection{Research Goal} % (fold)
\label{sub:research_goals}

The overarching goal of the proposed thesis is to form a theory of knowledge in software development, similar to the work of Reinhardt \textit{et al.} for knowledge workers~\cite{reinhardt2011knowledge,reinhardt2012awareness}. This theory emerges from our studies of the participatory process in software development. Figure~\ref{fig:knowledge_diagram} represents the knowledge theory in software development as we understand it now, and as this is a work in progress, the theory continues to develop and shape our research questions.

% subsection research_goals (end)

\subsection{Research Questions} % (fold)
\label{sub:research_questions}

First, we want to gain an understanding of the involved components (e.g., tools and channels, activities) in the knowledge theory. An important yet challenging task, as these components work together and investigating them separately produces only a partial understanding. Thus, our first research question is:
\begin{itemize}
    \item[\textbf{RQ1}] What characterizes a software development knowledge process?
    \begin{enumerate}
        \item[\textbf{A.}] What are the tools and channels developers use?
        \item[\textbf{B.}] What are differentiable roles in software development?
        \item[\textbf{C.}] What are typical knowledge activities and actions that developers perform?
    \end{enumerate}
\end{itemize}
Next, we aim to understand how the participatory process is affecting software development (enhances or hinders). The purpose is to use the components and terminology gained through RQ1. This leads to our second research question:
\begin{itemize}
    \item[\textbf{RQ2}] How does the participatory process shape and challenge software development?
    \begin{enumerate}
        \item[\textbf{A.}] How do communication channels shape development practices and activities?
        \item[\textbf{B.}] How do communication channels challenge developers?
    \end{enumerate}
\end{itemize}
And finally, we aim to demonstrate how a knowledge process, i.e., the participatory process, can be captured and investigated by using the resulting framework. This would serve as a validation of our work, and as a guide for other scholars. Thus, the third research question is:
\begin{itemize}
    \item[\textbf{RQ3}] How can we capture the participatory process?
\end{itemize}
Answering and continuously refining these research questions will guide our work.

% subsection research_questions (end)

\subsection{Expected Contributions} % (fold)
\label{sub:expected_contributions}
The proposed thesis aims to provide the following main contributions. First, we aim to characterize the participatory process in software development. This would help identify and define the involved components. Second, the proposed thesis will describe a \textit{Participatory Process} framework, that could be used by researchers and practitioners for studying or supporting software development. The framework will encourage researchers to not only use a narrow ``lens'' (e.g., communication channels, roles, or activities), but rather consider all aspects. And lastly, we aim to demonstrate the use of the proposed framework by applying it to a specific case study (to be designed later).
% subsection expected_contributions (end)

% section introduction (end)

%!TEX root = candidacy.tex

\section{Background} % (fold)
\label{sec:background}
We would like to start with a discussion on software development as a knowledge building process, describing the different types of knowledge (depicted in Fig.~\ref{fig:knowledge_diagram}). Then discuss communities of practice in software development as knowledge building communities and their role in the participatory culture. 

\subsection{Software Development as a Knowledge Process}
In 1985 Naur~\cite{naur1985programming} described programming as a \textit{``(knowledge) theory building''} activity, suggesting that programming (in the loose sense) is not just about producing a program, but rather are the developers' insights, theory, and knowledge built in the process. By describing two real cases, Naur shows that programming involves not only the source code and the documentation (i.e., \textbf{externalized knowledge}), but also the knowledge in developers' heads (i.e., \textbf{tacit knowledge}). He further explains that tacit knowledge is essential to software support and modification---if a developer was to leave the project, the tacit knowledge allowing these activities would be lost (e.g., design reasoning). This paper of course was written before developers formed and worked in communities of practice, where knowledge can be generated and maintained by the community, preserving the knowledge even when individual members leave. Tacit knowledge can be further decomposed into \textbf{procedural knowledge} and \textbf{declarative knowledge}~\cite{Robillard1999}. Procedural knowledge is dynamic and involves all the information related to the skills developed to interact with our environments. This knowledge is acquired from practice and experience. While declarative knowledge is based on facts, static, and concerned with the properties of objects, persons, and events and their relationships. Declarative knowledge is easy to describe and to communicate, as opposed to procedural knowledge.

The use of communication media in software development further extends the knowledge building process by enabling the transfer of knowledge between stakeholders. This facilitates individual learning and expression, as well as coordination and collaboration between members in a community. Wasko \textit{et al.}~\cite{Wasko2000} distinguish different theories of knowledge based on the kind of knowledge it helps capture or communicate:
\textbf{knowledge embedded in people} that may be tacit or embodied within people's heads, with its exchange typically done one-on-one or in small group interactions;
\textbf{knowledge as object} that exists in artifacts and can be accessed independently from any human being; and
\textbf{knowledge as public good} that is ``socially generated, maintained and exchanged within emergent communities of practice.'' We added a fourth type, based on our work~\cite{Storey2014}, designed to capture  \textbf{knowledge about people and social networks}.

\subsection{Communities of Practice (i.e., Knowledge Building Communities)}
Knowledge management in software development is an ongoing challenge~\cite{Bjørnson2008}; it involves the creation, capture, sharing and distribution of knowledge across a community. Since the formation and growth of a community relies on the complex interactions of its members~\cite{Storey2014}, it has been suggested that successful collaboration strongly depends on the social component of the community.

The core components of a \textbf{community of practice}~\cite{Lave1991,Wasko2000} are the \textbf{domain}, the \textbf{practice}, and the \textbf{community}~\cite{Wenger2011}. The domain, or shared interest, defines the identity of the community.
The practice identifies members of a community as \textbf{practitioners} that are constantly developing and sharing a set of resources (e.g., tools, documentation, histories, or experiences) to address recurring problems. And the community comprises the activities in which members engage in. Wenger~\cite{Wenger2011} asserts that, given the proper structure, practitioners can be the best option for managing the construction of knowledge (e.g., Stack Overflow). Communities of practice have been the focus of study in other domains as well. In the learning sciences, researchers refer to it as a \textbf{Knowledge Building Community} \textit{``that supports discourse and aims to advance the knowledge of the members collectively while still encouraging individual growth that will produce new experts and extend expertise within the community's domain''}.\footnote{\url{http://edutechwiki.unige.ch/en/Knowledge-building_community_model}}

Empowered by the \textit{``inexpensive and low barriers to publish, as well as the rapidly spreading peer-to-peer, large-scale communications made possible by social media''}, communities of practice have become an extensive part of software engineering~\cite{Storey2014}. The widespread adoption of socially enabled tools and channels facilitates virtual communities of practice and enables global software teams. Consequently, the availability of additional communication channels (e.g., micro-blogging) and a broad catalog of developer tools have increased the participatory culture among software developers. More importantly, the diversity of these tools and features posits a challenge in understanding which may be the best composition or combination of tools that will enhance the \textit{productivity} of collaborating developers---a challenge the proposed thesis aims to help with.

\subsection{A Note on Terminology}
This proposal follows the following rules regarding terminology:
In software engineering, a \textit{development process} (in some cases called a method or methodology) defines a set of development phases, activities and actions, roles, artifacts, and suggested tools and practices. For example, the agile process defines the actions (e.g., user research, design, coding, testing), the roles (e.g., integrator, product owner), and suggested practices (e.g., sprints, release early, daily stand-up meetings).
Based on the activity theory~\cite{tell2012activity} and the hierarchical model of human activity~\cite{engestrom1987learning}, we consider an \textit{activity} as high level set of actions where the subject is the developer. Where as an \textit{action} is a lower level entity, composed of \textit{operations}. Each of which is driven by a different need: motive, goal (or purpose), and condition~\cite{tell2012activity} (respectively). If we use Tell and Babar's provided example, an activity can be \textit{``designing''}, and is consisted of \textit{``analysis''}, \textit{``synthesis''}, and \textit{``evaluation''} actions. They further decompose actions into operations, e.g., \textit{``evaluation''} consists of \textit{``planing AE''} and \textit{``preparing and managing results''}. In this research proposal, we discuss the participatory process at the activity level, however, when defining the participatory process framework (later in the thesis phase), we'll follow Tell and Babar's activity theory, and consider the full activity system: subject, object, rules, instrument, community, and the division of labor.

% section background (end)
%!TEX root = candidacy.tex

\section{Understanding Communication Channels} % (fold)
\label{sec:what_we_learned}

In the following sections, we present what we know on the participatory process in software development and its main components (see Fig.~\ref{fig:knowledge_diagram})---communication channels, developers, and knowledge activities---thus answering \textbf{RQ1}. We highlight our work so far, and briefly describe the main findings. For each part, we indicate whether the work has been published, awaiting review, or partially completed. Here, we start with discussing how communication channels shape the participatory process in software development (\textbf{RQ1} and \textbf{RQ2}).

\subsection{Communication Media Shapes the Participatory Process\\ in Software Development} % (fold)
\label{ssub:fose2014}
\begin{description}
\item \textit{This work has been published in Proceedings of ICSE 2014, FOSE track}
\end{description}

In this phase, we examine how communication media supports the flow of knowledge in software development~\cite{Storey2014}. We form a retrospective of the different communication channels since the late 1960, and categorized them based on the theories of knowledge provided by Wasko \textit{et al.}~\cite{Wasko2000}.

\begin{figure}[h]
    \begin{center}
        \includegraphics[width=1.0\textwidth]{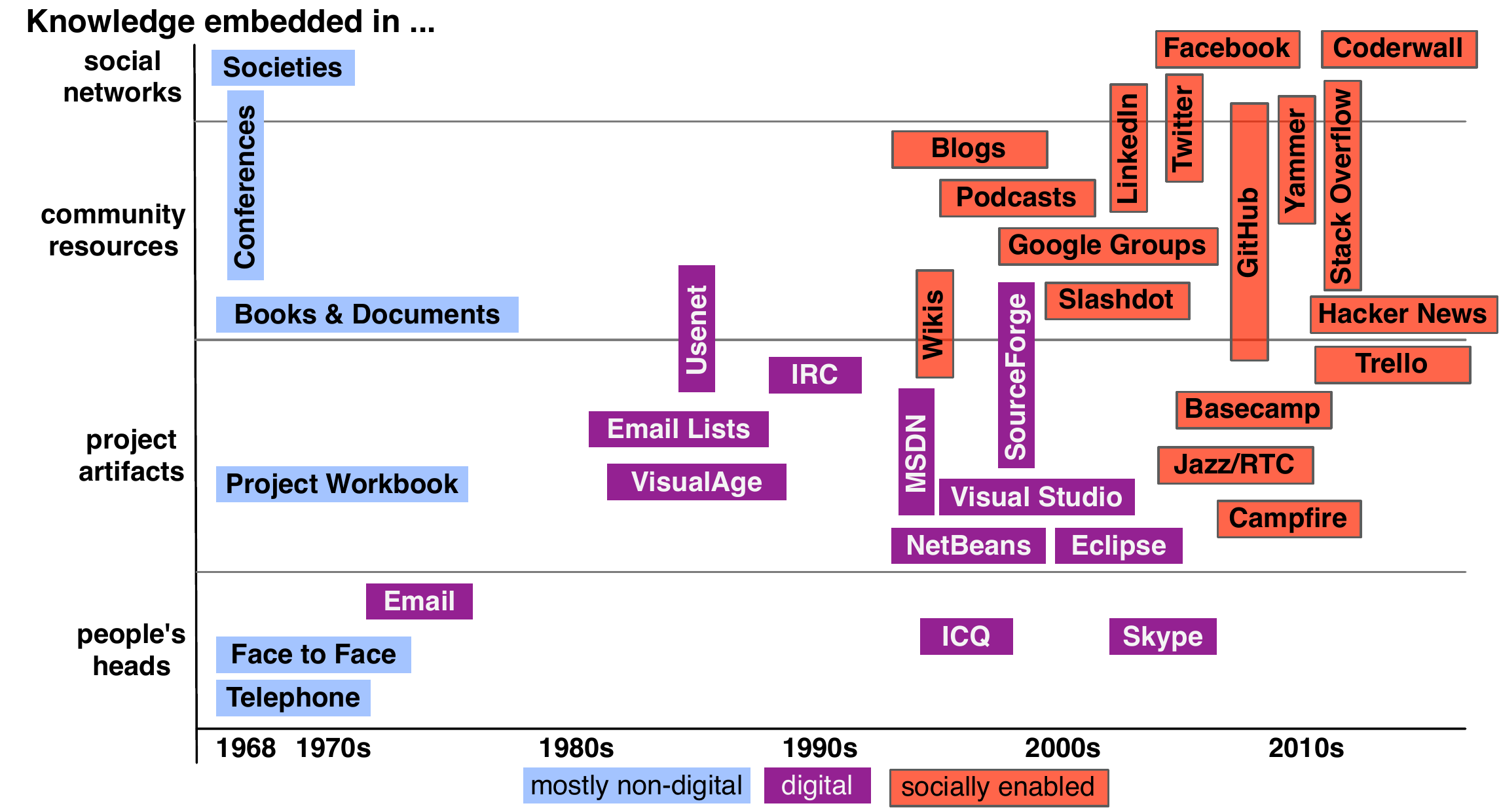}
        \caption{Media channels over time and how they support the transfer of developer knowledge. Note that some channels overlap multiple types of knowledge communication.}
        \label{fig:timeline}
    \end{center}
\end{figure}

Our findings reveal the transition has happened over time, showing the emergence of channels that are increasingly socially enabled, and a rise in the adoption of channels for exchanging community knowledge and social networking content (as illustrated in Fig.~\ref{fig:timeline}). The timeline highlights how we have entered a social era in terms of media use in software engineering. Although this provides us some insightful research findings about the implications of the paradigm shift in software engineering, we lack insights on how various media channels are used in combination to support software engineering activities.

To try to understand how the newer generation of developers uses a broad range of social media channels, we conducted a large scale survey (1,516 responses) with members of a thriving participatory community through GitHub. The preliminary findings indicate that our respondents use a median of 12 unique channels (mean: 11.55, min: 1, max: 21). While 25\% of the respondents use between 14 and 21 unique channels!
The survey also unveils an initial set of challenges developers face in this new participatory culture---most mentioned challenges were: media literacy, anti-social behavior, trusting content, and vendor lock-in. We used this initial set of challenges as a starting point for a follow-up study (described in Section~\ref{sub:dev_challenges}) where we investigated how communication media shapes and challenges a participatory culture in software development~\cite{Storey2015}.

\subsubsection{How Does This Fit in The Thesis}
This study allows us to gain a retrospective of the tools and channels used to support software development, and understand their evolution (addressing \textbf{RQ1-A}). Moreover, this study lays the basis and served as a precursor for studying the challenges developers face (contributing to \textbf{RQ2-B}), and the activities they do (contributing to \textbf{RQ1-C})---both are critical to understanding the knowledge building process.

% subsection fose2014 (end)

\subsection{Communication Channels Shape Developer Practices:\\Understanding How Software Teams Use Slack} % (fold)
\label{sub:cscw2016_poster}
\begin{description}
\item \textit{This work is under review for CSCW 2016, Interactive Poster Submission}
\end{description}

\begin{chapquote}{Marshall McLuhan}
\vspace{12pt}
``We shape our tools and thereafter our tools shape us.''
\end{chapquote}

Slack is playing an increasingly significant role in shaping the way software developers collaborate and communicate. In just two years, it has gained more than 1.1 million daily users and over 900k integrations\footnote{\url{http://techcrunch.com/2015/06/24/as-slack-hits-1m-daily-users-and-900k-integration-installs-it-hires-april-underwood-as-head-of-platform/}}. Slack not only facilitates team messaging and archiving, it also supports a wide plethora of integrations to external services and bots (e.g., GitHub, Asana, Jira, Hubot). Moreover, Slack is disrupting and shaping developers' activities and practices within the modern, social, fast-moving and sometimes overwhelming development environment~\cite{Storey2014}.

\begin{chapquote}{an anonymous developer}
\vspace{11pt}
``If Slack is taken out, it's just like turning our lights off. We're blind.''
\end{chapquote}

To understand how Slack impacts development team dynamics, we designed an exploratory study to investigate how developers use Slack and how they benefit from it. Understanding \textbf{how} and \textbf{why} developers use Slack to support their work is essential to gaining insights into the modern software development activities, and how these activities are being adapted to the participatory process.

\subsubsection{Methodology}
In this exploratory study, our goal is to understand how Slack and its integrations are used by software developers. In turn, allowing us to understand how modern communication tools are shaping development practices and activities (\textbf{RQ2-A} and contributing to \textbf{RQ1-A}). Our study is guided by the following secondary research questions:
\begin{enumerate}
    \item[\textbf{SRQ1}] What do developers use Slack for and how do they benefit?
    \item[\textbf{SRQ2}] What bots do developers use and why do they use them?
\end{enumerate}

So far, we have deployed two surveys with open-ended questions to developers who adopted Slack. For the \textbf{first survey}, we targeted software developers who use Slack. We promoted it to 30 development-related Slack teams (that are publicly open) and also promoted it through Twitter. We received 53 responses to the first survey\footnote{Survey used in the first phase: \url{http://goo.gl/forms/mnGhSZCNtY}}.
From the analysis of this survey, we realized that bots played a significant role in software development processes.  Thus, for a \textbf{second iteration of the survey}, we refined some of the questions\footnote{Survey for the second phase: \url{http://goo.gl/forms/lZpGXPE6kH}} and focused the distribution of it to developers that customized Slack bots---specifically we emailed 650 developers who forked or starred the ``hubot-slack'' project on GitHub\footnote{\url{https://github.com/slackhq/hubot-slack}}. We received 51 responses to this second survey.

\subsubsection{What Do Developers Use Slack For and How Do They Benefit?}
Our analysis reveals that developers use Slack for: personal, team-wide, and community-wide reasons.

\textbf{Personal benefits:}
Developers use Slack as a gateway to \textbf{discover and aggregate news and information.} \textit{``[I use slack for] RSS reader/bookmarking site for reliably interesting/relevant blogs (e.g. Signal vs. Noise, Rocky Mountain Institute, etc)"}.  They also use the instant messaging features to support
\textbf{networking and social activities} with developers who share similar interests (e.g., Android developers) or have similar jobs.
Surprisingly, they also use it for \textbf{fun}, as participants told us they use it for \textit{``sharing gifs and memes''}
% could cut the mention of Giphy here as it may appear in RQ2)
(through Giphy, one of the most popular bots used) and for \textit{``gaming''}.

\textbf{Team-wide purposes:}
Slack's messaging feature is used widely for \textbf{communication}:
\textit{``[I use Slack for] communication with teammates (almost exclusively, we're a remote team).''}
But, the way it is used varies. For some, it is used for remote meetings and note taking, for communication with other stakeholders (such as customers through live-chats) and for non-work topics. It further supports \textbf{team collaboration} through team management, file and code sharing, development operation notifications, and software deployments (i.e., \textbf{Dev-Ops}) and team Q\&A.

\textbf{Community support:} Slack provides extensive support for \textbf{participation in communities of practice}, or special interest groups: \textit{``[I use Slack for] keeping up with specific frameworks/communities''}. Other participants mentioned using it for \textit{``bouncing ideas off of other people in the community''} or use it for \textit{``learning about new tools and frameworks for developing applications.''}

\subsubsection{How Does This Fit in The Thesis}
Slack is the perfect motivation for building a knowledge theory, or a participatory process framework. This is a new tool/channel which is extensively used by software teams. However, we lack the tools and terminology to understand \textit{why} and \textit{how} it is used. Furthermore, Slack shapes and changes the behavior of developers (\textbf{RQ2-A}), while those in turn shape the design of the tool itself. To follow McLuhan's tetrad~\cite{mcluhan1967medium}: How does Slack \textbf{enhance} and support software development teams (e.g., collaboration and awareness)? What does it \textbf{make obsolete} (e.g., data is no longer siloed) ? And how does it flip when \textbf{taken to extreme} (e.g., what happens when Slack is down)? This thesis could help in answering those questions (\textbf{RQ2-B}).

\subsubsection{What is Next?}

This exploratory study was a preliminary phase. It allowed us a glimpse into \textbf{how} and \textbf{why} developers use Slack. However, our insights are somewhat superficial, as expected for an exploratory phase. Thus, the next step should aim for extracting deeper insights for the same research questions, e.g. through the use of interviews with developers. In fact, we have already moved forward by conducting an interview with developers from a local company (SendWithUs) that uses Slack, revealing interesting insights. But more importantly, we lack an appropriate framework or typology of knowledge activities in software development, and as a result it limited our ability in refining our research method and findings (e.g., frame the findings in terms of supported activities). We are currently working on building a \textbf{knowledge activity typology} (further described in Section~\ref{sub:developer_activities}). Additionally, we need to look at what happens when this new medium is taken to the extreme. What challenges does this new medium introduce? For example, some communities already object to using Slack\footnote{\url{https://drewdevault.com/2015/11/01/Please-stop-using-slack.html}}.

\vspace{5mm}
As part of our study on how Slack is used by development teams, we noticed an interesting phenomena of \textit{social bots} that support developers (and other stakeholders). Next, we discuss the natural extension from the \textit{social programmer} to the use of \textit{social bots} in supporting software development, and how it may affect knowledge flow.

% subsection cscw2016_poster (end)

\subsection{The Rise of The Social Bots in Software Development} % (fold)
\label{sub:bots}
\begin{description}
\item \textit{This work has been partially completed}
\end{description}
Social media has dramatically changed the landscape of software engineering, challenging some old assumptions about how developers learn and work with one another. Bringing the rise of the \emph{social programmer} who actively participates in online communities and openly contributes to the creation of a large body of \emph{crowdsourced socio-technical content}~\cite{Storey2014}. This social developer is a \textbf{lifelong learner}, engaged in a continuous cycle of updating one's skills and further developing one's competencies in both formal and informal educational settings~\cite{reinhardt2011knowledge}.

Developers engaging in the participatory culture have recently adopted a new and versatile channel---the \textbf{social media bots}\footnote{\url{http://www.wired.com/2015/10/the-most-important-startups-hardest-worker-isnt-a-person/}}. Our exploratory study of how and why developers use Slack (see Section~\ref{sub:cscw2016_poster}), revealed that developers use bots to retrieve or post messages (e.g., the Twitter bot reposts tweets to Slack), to integrate other communication channels such as audio, video, screen sharing (e.g., Screenhero), and email. Bots are also used for information acquisition through news aggregators. Some developers customize bots to fit their needs (e.g., to initiate coordination in the team, or to support trivial tasks). However, not much is known on how these bots affect software development and the knowledge building process.

\subsubsection{How Does This Fit in The Thesis}
In a similar manner to studying Slack, studying how social bots shape (or hinder) software development can guide researchers and practitioners (\textbf{RQ2-A}). Moreover, bots could perhaps become a separate entity in the knowledge theory diagram (see Fig.~\ref{fig:knowledge_diagram}).

\subsubsection{What is Next?}
The use of social bots in software development is an emerging phenomena, one that has not been studied so far. The use of bots changes and shapes the developers' practices and the activities they partake in. For example, developers interact with bots inside the communication channels (e.g., Slack), transforming these communication channels into a command-line interface. As a result, introducing benefits such as improved team awareness, and supporting the ability for non-technical stakeholder to interact directly with the system (e.g., deploying a new version through the use of a bot). However, what challenges does this introduce? (e.g., signal vs. noise issue).
%Another aspect to be examined is how bots support and shape the company culture. There is early evidence for bots being used for building and promoting the company culture---but what happens when the company's culture changes?
Thus, the following research questions should guide us in the next phase:
\begin{enumerate}
    \item[\textbf{SRQ1}] What knowledge activities developers use bots for? (contributing to \textbf{RQ1-C})
    \item[\textbf{SRQ2}] What challenges does the use of social bots introduces? (contributing to \textbf{RQ2-B})
    \item[\textbf{SRQ3}] How do bots affect the knowledge flow in software development?
\end{enumerate}

As an initial phase, we have conducted an interview with developers in a local company, SendWithUs. The interview is being currently transcribed.

% subsection bots (end)

% section social_dev (end)
%!TEX root = candidacy.tex
\newpage
\section{Understanding Developers and Their Roles} % (fold)
\label{sec:stakeholders}
In this section, we present our work relating to the study of developers. We focus in this section on the human aspect of the participatory process, that is, we discuss studies aiming to understand the challenges developers face (\textbf{RQ2-B}), the way they collaborate (\textbf{RQ1-A}), and the roles they fulfill (\textbf{RQ1-B}). First, we start by examining challenges developers experience when using communication media to support software development.

\subsection{How Communication Channels Challenge A Participatory Culture\\in Software Development} % (fold)
\label{sub:dev_challenges}
\begin{description}
\item \textit{This work is under review for a TSE journal}
\end{description}

\begin{chapquote}{Marshall McLuhan}
\vspace{12pt}
``What does the medium flip into when pushed to extremes?''
\end{chapquote}

Software developers use many different communication tools and channels in their work. The diversity of these tools has dramatically increased over the past decade, giving rise to a wide range of socially-enabled communication channels and social media that developers use to support their activities. The availability of such social tools is leading to a participatory culture of software development, where developers want to engage with, learn from, and co-create software with other developers. However, the interplay of these social channels, as well as the opportunities and \textbf{challenges} they may create when used together within this participatory development culture, are not yet well understood. In this study, we investigated how the choice of communication channels shapes the activities that developers partake in within a participatory culture of development, as well as explore the challenges they may face.

\subsubsection{Methodology}
We designed and conducted a survey to learn how developers use tools to support their knowledge activities, to learn which media channels are important to them, and to determine which challenges they may face. Our secondary research questions were:
\begin{itemize}
    \item[\textbf{SRQ1}] \textit{Who is the social programmer} that participates in these communities?  
    % That is, what are the demographics of developers using GitHub?  
    \item [\textbf{SRQ2}] \textit{Which communication channels} do these developers use to support development activities?
    \item [\textbf{SRQ3}] Which communication channels are the most \textit{important} to developers and \textit{why}?
    \item [\textbf{SRQ4}] \textit{What challenges} do developers face using an ecosystem of communication channels to support their activities?
\end{itemize}

We deployed the same survey during two different time periods: at the end of 2013 and at the end of 2014, and 1,516 developers responded (21\% response rate).

\subsubsection{The Challenges Developers Face Using Communication Channels}

\begin{figure}[h]
    \begin{center}
        \includegraphics[width=0.6\columnwidth]{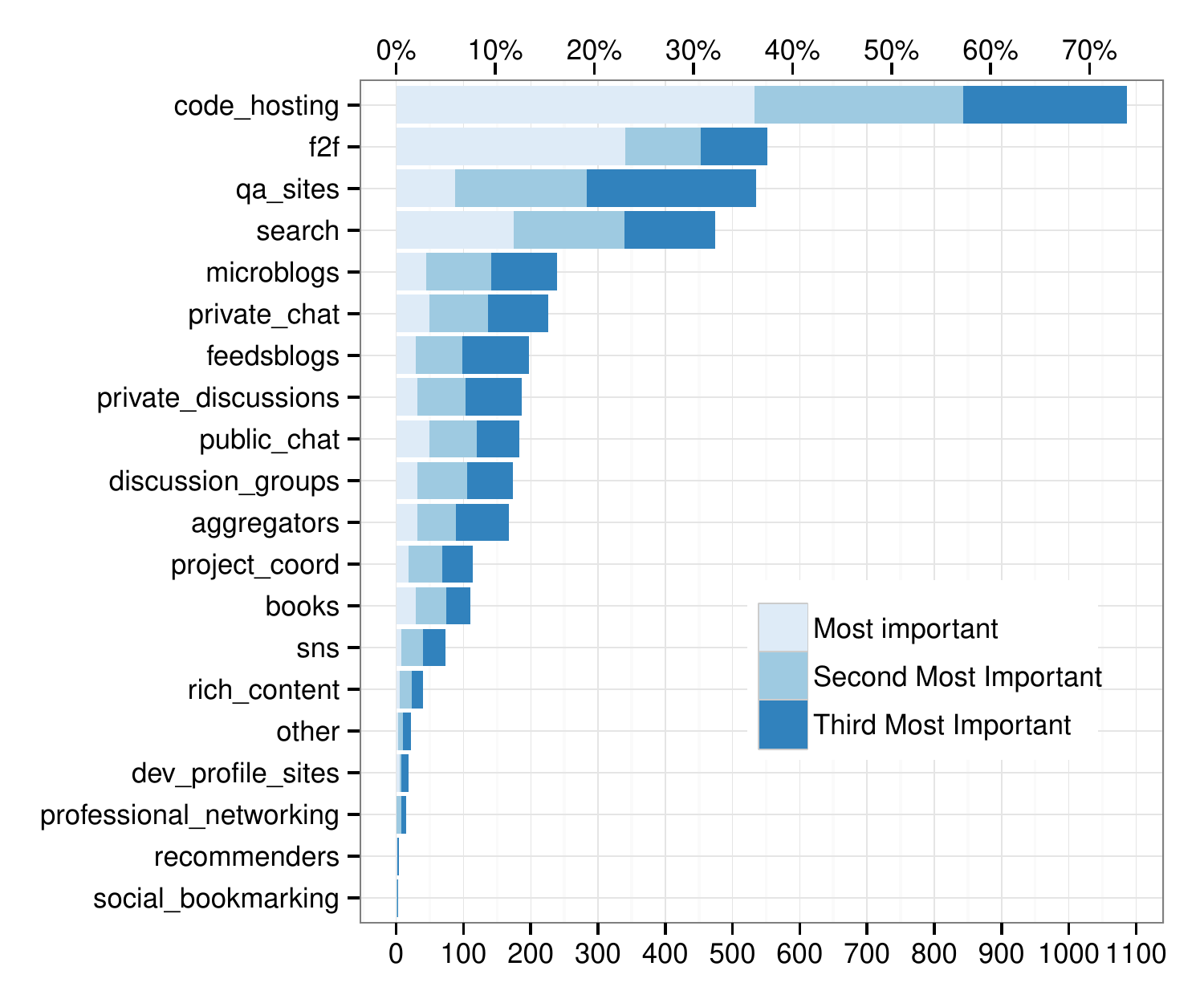}\\
        \caption{Number of responses per channel indicating the importance of each channel.}
        \label{fig:number_of_responses_per_channel}
    \vspace{-3pt}
    \end{center}
\end{figure}

We found that on average developers indicated they use 11.7 channels across all activities, with a median of 12. We also asked developers to indicate the top three channels that are important to them (Figure~\ref{fig:number_of_responses_per_channel} shows the number of responses given per channel). Surprised by the high number of channels, we asked ourselves what happens \textit{``when the medium is pushed to the extreme?''}

\begin{figure}[h]
    \begin{center}
         \vspace{-13pt}
        \includegraphics[width=0.5\columnwidth]{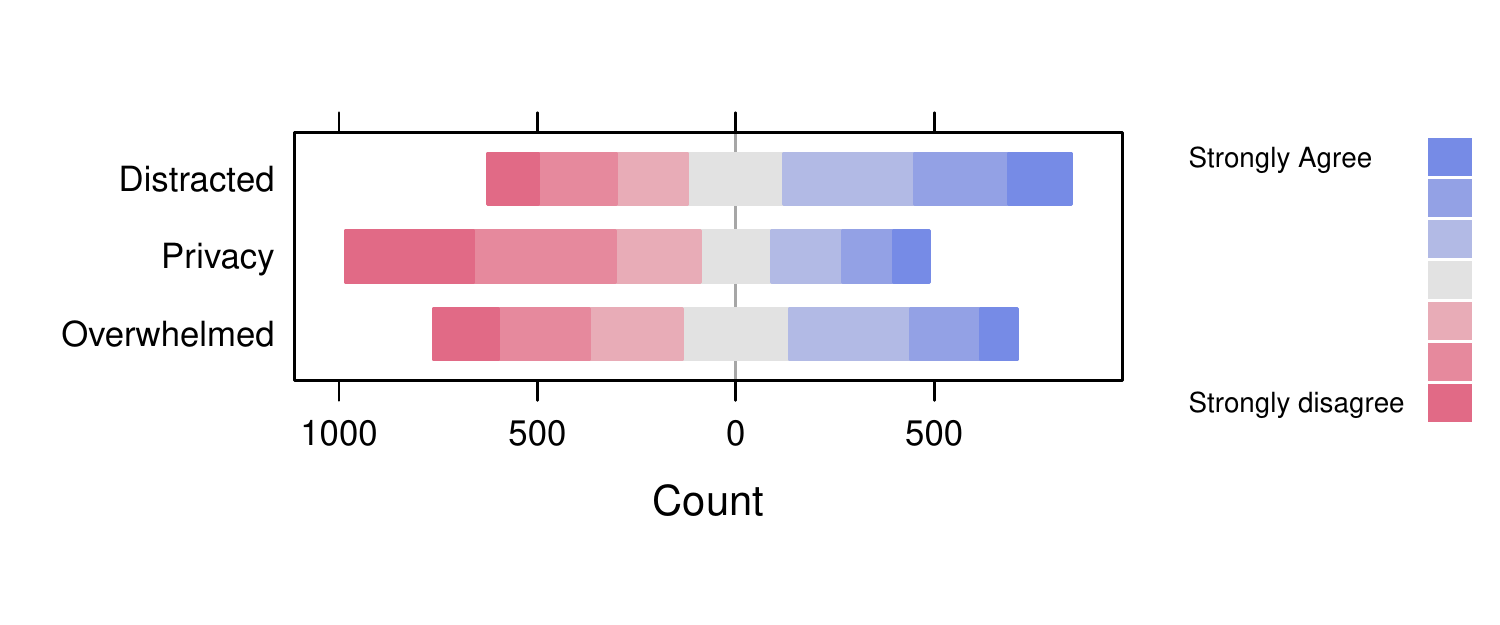}
         \vspace{-26pt}
        \caption{Frequency of responses to Likert questions probing on developer challenges.}
        \label{fig:challengesLikert}
    \end{center}
\end{figure}

Thus, we would like to focus on research question \textbf{SRQ4} which inquires about the challenges developers face using communication channels. From our previous work \cite{singer2013mutual,Singer2014,Storey2014}, we were already aware that developers face challenges related to distractions, privacy and being overwhelmed by communication chatter using social media channels. Consequently, the survey asked specifically whether the respondents experienced these concerns. Fig.~\ref{fig:challengesLikert} shows the results from the Likert-style questions of the survey. We note that privacy is not a big concern for everyone, whereas being interrupted and feeling overwhelmed by communication traffic are issues for more developers.

We further asked the respondents to share with us any \textbf{additional challenges} they face through an open-ended text question. 432 respondents (356 to the 2013 survey, 76 to the 2014 deployment) either elaborated on the challenges mentioned above, or informed us about other challenges they face. A wide variety of challenges were reported and we coded, sorted, clustered and then categorized them using an open coding and iterative clustering technique. Postman, who studied the use of media in communities of practice extensively, refers to a \textit{media ecology} and suggests we undertake the study of ``the entire environments, their structure, content and \textbf{impact on people}.''~\cite{Postman1971}. We note that the categories we found mirror the main areas of study suggested by Postman. The main categories of challenges that emerged from our analysis are as follows (we report these challenges in detail in the paper itself):  
\begin{itemize}[noitemsep]
\item Developer issues
\item Collaboration and coordination hurdles
\item Barriers to community building and participation
\item Social and human communication challenges
\item Communication channel affordances, literacy and friction
\item Content and knowledge management concerns
\end{itemize}

\subsubsection{How Does This Fit in The Thesis}
Understanding developers and stakeholder roles is an essential part of building a knowledge theory and understanding the participatory process. Thus, the study of the challenges developers face was important. In fact, preliminary findings regarding the challenges relating to the developer activities shows similar challenge categories to the ones emerging from this study.

\subsubsection{What is Next?}
The challenges in this study focused on challenges when using communication channels to support software development. However, there is a need to also understand the challenges developers have in regard to the knowledge activities they partake in (something we already began investigating, and is discussed in Section~\ref{sub:developer_activities}).

% subsection dev_challenges (end)

\subsection{Developer Roles in Software Development} % (fold)
\label{sub:developer_roles_in_software_development}
\begin{description}
\item \textit{This work has not been done yet}
\end{description}
% subsection developer_roles_in_software_development (end)
\subsubsection{How Does This Fit in The Thesis}
Developer (stakeholder) roles is an essential part of understanding the knowledge flow, and it would allow to identify different types of stakeholders and better understand the ``expected behavior pattern'' (\textbf{RQ1-B}).

\subsubsection{What is Next?}
This study hasn't been conducted yet. A main reason was that we wanted first to build the activity typology to guide us in this phase. The proposed methodology would be a mixed-methods methodology, consisting of a survey, interviews, and inspection of team data (e.g., Slack conversations, GitHub). The expected result is a typology of roles, describing the: roles, a description of each role, and typical knowledge activities and channels/tools associated to that role.

\subsection{Studying Collaboration As Part of The Participatory Process} % (fold)
\label{sec:collaboration}
\begin{description}
\item \textit{This work has been published as a preliminary work at CHASE 2015 workshop,\\and is under review for ICSE 2016, research track}
\end{description}

Comprehending the synergetic nature of software development is difficult as collaboration is a complex activity that relies on awareness, communication, and coordination. However, gaining an understanding of collaboration in today's participatory development culture is even more challenging as there is a growing number of stakeholders involved in its development and feature needs evolve rapidly as software is continuously deployed. In our experience, the existing models of collaboration for software development (e.g., the 3C model~\cite{Fuks2005}) are not powerful enough to capture the way different actors and stakeholders work together, nor do they capture how stakeholders have to respond to emergent requirements. These models focus heavily on task coordination, task completion, and stakeholder communication, but they  do not adequately take into consideration knowledge co-construction by emergent stakeholders.

The challenge of modeling dynamic collaboration processes is not unique to software engineering and is shared by other domains that focus on knowledge work. To address the above mentioned shortcomings of the existing collaboration models, we worked on borrowing, adapting, and extending a model of regulated learning from the education domain known as the Model of Regulation~\cite{Hadwin2011}. \textit{The Model of Regulation} describes distinct \textbf{regulation forms} (self-,co-, and socially shared-regulation), it defines \textbf{regulation processes} (task understanding, goal setting, enacting, large scale adaptation, and monitoring \& evaluating), and it defines the \textbf{types of process support} provided by tools or channels (structuring support, mirroring support, awareness tools, and guiding systems). In this work, we aim to explore the viability of using the Model of Regulation to understand collaboration support in software development (e.g., in tools such as GitHub, Slack, Trello).

\subsubsection{Methodology}
Our research methodology consisted of a two-phase qualitative case study~\cite{runeson2009,yin2013case} investigation. In the first phase, our purpose was \emph{exploratory} as we investigated the feasibility of using the Model of Regulation as a way to understand the dynamics of the collaboration involved in the development of an open source project. The research question that guided this phase is:
\begin{itemize}[noitemsep]
    \item[\textbf{SRQ1}] Can the Model of Regulation be used to identify how regulation occurs in software development?
\end{itemize}

Phase I revealed new complexities in the Model of Regulation when multiple actor types with different investments---\textit{Users} and \textit{Contributors}---are involved in collaboration.
Having adapted and refined the model, we then transitioned into phase II and investigated how regulation occurs in software development and how users participate in that regulation. We were guided by our second research question:
\begin{itemize}[noitemsep]
    \item[\textbf{SRQ2}] How do users participate in the regulation of an open source project?
\end{itemize}
Both phases of our study investigated the Neo4J\footnote{\url{http://neo4j.com/open-source-project}} community. It is important to note that despite using the same \textit{``case''} for both phases, they had different purposes (\textit{exploratory} vs. \textit{descriptive}) and relied on different processes and data.

\subsubsection{Understanding The User's Role in the Regulation of an Open Source Project}
By adapting and applying the Model of Regulation to a software project, we were able to gain an understanding in the user's role in the regulation of the project. Our findings show that: user participation \textbf{promotes task understanding} processes among contributors; user participation \textbf{contributes to the definition of project goals and standards}; close collaboration between users and contributors \textbf{helps them reach agreement} on the strategic selection of tools and resources to achieve their goals; and, users \textbf{play an important role in evaluating the project and supporting collaboration};

\begin{figure}[h!]
    \begin{center}
        \includegraphics[width=0.5\textwidth]{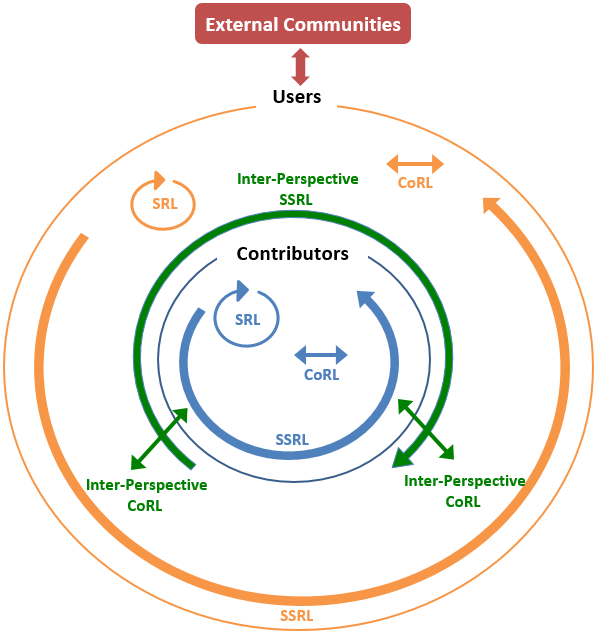}
        \caption{A representation of regulation forms in an open source project}
        \label{fig:regForms}
    \end{center}
\end{figure}

These findings may not seem surprising, however, the main contribution of this work allows to identify and capture these instances of regulation and collaboration (examples and an extensive description of the Model of Regulation are described in detail in the paper itself). Figure~\ref{fig:regForms} shows a representation of regulation process in an open source project (taken in a snapshot in time).

\subsubsection{How Does This Fit in The Thesis}
The adapted Model of Regulation acts as a lens that allowed us to understand the specific types of regulation between users and project contributors in a collaborative software development project, a project that is an example of a participatory community. Furthermore, the Neo4J community showcases several knowledge activities, e.g., \textit{co-authoring}, which can be supported by regulation process and thus allowing us to better understand these collaborative activities as part of the participatory process.

\subsubsection{What is Next?}
Next step should consider an analysis of the contributor's community and their use of communication tools to support their regulation processes.
In this study, we studied communication in three different channels (namely GitHub issues, Google Groups and Stack Overflow) as a way to triangulate how regulation forms and processes from the Model of Regulation may be evident in collaborative software development. Future work should delve more deeply into the question of channel affordances for regulation, as well as consider other channels such as Slack (which the Neo4J group have recently started to use\footnote{\url{http://neo4j.com/blog/public-neo4j-users-slack-group}}). These insights about regulation and communication channel affordances should help practitioners decide which constellation of tools would give the best support for regulation in their collaborative projects.
% section stakeholders (end)
%!TEX root = candidacy.tex
\newpage
\section{Understanding Knowledge Activities in Software Development} % (fold)
\label{sec:activity_theory}

Next, we present our work relating to the study of knowledge activities in software development (\textbf{RQ1-C}). This work is a \textit{cornerstone} in my thesis. We describe below our gradual progress achieved over several studies.

\definecolor{c00}{gray}{1.0}
\definecolor{c10}{gray}{0.9}
\definecolor{c20}{gray}{0.8}
\definecolor{c30}{gray}{0.7}
\definecolor{c40}{gray}{0.6}
\definecolor{c50}{gray}{0.5}
\definecolor{c60}{gray}{0.4}
\definecolor{c70}{gray}{0.3}
\definecolor{c80}{gray}{0.2}
\definecolor{c90}{gray}{0.1}

\definecolor{c_analog}{rgb}{0.64,0.75,1.0}
\definecolor{c_digital}{rgb}{0.64,0.0,0.57}
\definecolor{c_social}{rgb}{1.0,0.29,0.29}

\begin{table}[b!]
    \begin{center}
        \renewcommand{\arraystretch}{1.1}
        \begin{tabular}{r!{\vrule width 2pt}l|l!{\vrule width 2pt}l|l|l|l|l|l|l!{\vrule width 2pt}l|l|l|l|l|l|l|l|l|l!{\vrule width 2pt}}
            ~ & \begin{sideways}Face-to-face\end{sideways} & \begin{sideways}Books\end{sideways} & \begin{sideways}Web Search\end{sideways} & \begin{sideways}Content Recommenders\end{sideways} & \begin{sideways}Rich Content\end{sideways} & \begin{sideways}Private Discussions\end{sideways} & \begin{sideways}Discussion Groups\end{sideways} & \begin{sideways}Public Chat\end{sideways} & \begin{sideways}Private Chat\end{sideways} & \begin{sideways}Feeds and Blogs\end{sideways} & \begin{sideways}News Aggregators\end{sideways} & \begin{sideways}Social Bookmarking\end{sideways} & \begin{sideways}Q\&A Sites\end{sideways} & \begin{sideways}Prof. Networking Sites\end{sideways} & \begin{sideways}Developer Profile Sites\end{sideways} & \begin{sideways}Social Network Sites\end{sideways} & \begin{sideways}Microblogs\end{sideways} & \begin{sideways}Code Hosting Sites\end{sideways} & \begin{sideways}Project Coordination Tools\end{sideways}\\
\noalign{\hrule height 2pt}
            ~ & \multicolumn{2}{c!{\vrule width 2pt}}{\cellcolor{c_analog!75}analog} & \multicolumn{7}{c!{\vrule width 2pt}}{\cellcolor{c_digital!75}digital} & \multicolumn{10}{c!{\vrule width 2pt}}{\cellcolor{c_social!75}digital and socially enabled} \\
\noalign{\hrule height 2pt}
% start of generated latex (`cd task-channel-table && ./devsurvey.rb devsurvey.csv`)
Stay Up to Date & \cellcolor{c50}~ & \cellcolor{c30}~ & \cellcolor{c70}~ & \cellcolor{c00}~ & \cellcolor{c20}~ & \cellcolor{c30}~ & \cellcolor{c40}~ & \cellcolor{c20}~ & \cellcolor{c30}~ & \cellcolor{c50}~ & \cellcolor{c50}~ & \cellcolor{c00}~ & \cellcolor{c50}~ & \cellcolor{c10}~ & \cellcolor{c10}~ & \cellcolor{c30}~ & \cellcolor{c30}~ & \cellcolor{c70}~ & \cellcolor{c00}~ \\
\hline
Find Answers & \cellcolor{c40}~ & \cellcolor{c10}~ & \cellcolor{c80}~ & \cellcolor{c00}~ & \cellcolor{c00}~ & \cellcolor{c20}~ & \cellcolor{c40}~ & \cellcolor{c20}~ & \cellcolor{c20}~ & \cellcolor{c20}~ & \cellcolor{c00}~ & \cellcolor{c00}~ & \cellcolor{c80}~ & \cellcolor{c00}~ & \cellcolor{c00}~ & \cellcolor{c00}~ & \cellcolor{c10}~ & \cellcolor{c40}~ & \cellcolor{c00}~ \\
\hline
Learn & \cellcolor{c40}~ & \cellcolor{c40}~ & \cellcolor{c60}~ & \cellcolor{c00}~ & \cellcolor{c20}~ & \cellcolor{c20}~ & \cellcolor{c30}~ & \cellcolor{c20}~ & \cellcolor{c20}~ & \cellcolor{c40}~ & \cellcolor{c20}~ & \cellcolor{c00}~ & \cellcolor{c60}~ & \cellcolor{c00}~ & \cellcolor{c10}~ & \cellcolor{c10}~ & \cellcolor{c20}~ & \cellcolor{c60}~ & \cellcolor{c00}~ \\
\hline
Discover Others & \cellcolor{c30}~ & \cellcolor{c00}~ & \cellcolor{c30}~ & \cellcolor{c00}~ & \cellcolor{c10}~ & \cellcolor{c10}~ & \cellcolor{c20}~ & \cellcolor{c10}~ & \cellcolor{c10}~ & \cellcolor{c30}~ & \cellcolor{c20}~ & \cellcolor{c00}~ & \cellcolor{c30}~ & \cellcolor{c10}~ & \cellcolor{c10}~ & \cellcolor{c20}~ & \cellcolor{c30}~ & \cellcolor{c60}~ & \cellcolor{c00}~ \\
\hline
Connect With Others & \cellcolor{c30}~ & \cellcolor{c00}~ & \cellcolor{c00}~ & \cellcolor{c00}~ & \cellcolor{c00}~ & \cellcolor{c30}~ & \cellcolor{c20}~ & \cellcolor{c20}~ & \cellcolor{c20}~ & \cellcolor{c10}~ & \cellcolor{c00}~ & \cellcolor{c00}~ & \cellcolor{c20}~ & \cellcolor{c10}~ & \cellcolor{c00}~ & \cellcolor{c20}~ & \cellcolor{c30}~ & \cellcolor{c60}~ & \cellcolor{c00}~ \\
\hline
Get and Give Feedback & \cellcolor{c30}~ & \cellcolor{c00}~ & \cellcolor{c00}~ & \cellcolor{c00}~ & \cellcolor{c00}~ & \cellcolor{c30}~ & \cellcolor{c30}~ & \cellcolor{c20}~ & \cellcolor{c20}~ & \cellcolor{c10}~ & \cellcolor{c00}~ & \cellcolor{c00}~ & \cellcolor{c30}~ & \cellcolor{c00}~ & \cellcolor{c00}~ & \cellcolor{c10}~ & \cellcolor{c20}~ & \cellcolor{c60}~ & \cellcolor{c10}~ \\
\hline
Publish Activities & \cellcolor{c10}~ & \cellcolor{c00}~ & \cellcolor{c00}~ & \cellcolor{c00}~ & \cellcolor{c00}~ & \cellcolor{c10}~ & \cellcolor{c10}~ & \cellcolor{c00}~ & \cellcolor{c10}~ & \cellcolor{c20}~ & \cellcolor{c00}~ & \cellcolor{c00}~ & \cellcolor{c10}~ & \cellcolor{c10}~ & \cellcolor{c00}~ & \cellcolor{c20}~ & \cellcolor{c30}~ & \cellcolor{c80}~ & \cellcolor{c00}~ \\
\hline
Watch Activities & \cellcolor{c10}~ & \cellcolor{c00}~ & \cellcolor{c10}~ & \cellcolor{c00}~ & \cellcolor{c00}~ & \cellcolor{c00}~ & \cellcolor{c10}~ & \cellcolor{c10}~ & \cellcolor{c00}~ & \cellcolor{c30}~ & \cellcolor{c10}~ & \cellcolor{c00}~ & \cellcolor{c10}~ & \cellcolor{c00}~ & \cellcolor{c00}~ & \cellcolor{c20}~ & \cellcolor{c30}~ & \cellcolor{c80}~ & \cellcolor{c00}~ \\
\hline
Display Skills/Accomplishments & \cellcolor{c10}~ & \cellcolor{c00}~ & \cellcolor{c00}~ & \cellcolor{c00}~ & \cellcolor{c00}~ & \cellcolor{c00}~ & \cellcolor{c00}~ & \cellcolor{c00}~ & \cellcolor{c00}~ & \cellcolor{c20}~ & \cellcolor{c00}~ & \cellcolor{c00}~ & \cellcolor{c10}~ & \cellcolor{c30}~ & \cellcolor{c10}~ & \cellcolor{c20}~ & \cellcolor{c20}~ & \cellcolor{c60}~ & \cellcolor{c00}~ \\
\hline
Assess Others & \cellcolor{c20}~ & \cellcolor{c00}~ & \cellcolor{c10}~ & \cellcolor{c00}~ & \cellcolor{c00}~ & \cellcolor{c10}~ & \cellcolor{c10}~ & \cellcolor{c10}~ & \cellcolor{c10}~ & \cellcolor{c20}~ & \cellcolor{c00}~ & \cellcolor{c00}~ & \cellcolor{c20}~ & \cellcolor{c10}~ & \cellcolor{c00}~ & \cellcolor{c10}~ & \cellcolor{c20}~ & \cellcolor{c70}~ & \cellcolor{c00}~ \\
\hline
Coordinate With Others & \cellcolor{c30}~ & \cellcolor{c00}~ & \cellcolor{c00}~ & \cellcolor{c00}~ & \cellcolor{c00}~ & \cellcolor{c50}~ & \cellcolor{c20}~ & \cellcolor{c20}~ & \cellcolor{c40}~ & \cellcolor{c00}~ & \cellcolor{c00}~ & \cellcolor{c00}~ & \cellcolor{c00}~ & \cellcolor{c00}~ & \cellcolor{c00}~ & \cellcolor{c00}~ & \cellcolor{c00}~ & \cellcolor{c70}~ & \cellcolor{c30}~ \\
\hline
% end of generated latex (`cd task-channel-table && ./devsurvey.rb devsurvey.csv`)
        \end{tabular}
        \\
        \vspace{0pt}
        \begin{tabular}{ll|l|l|l|l|l|l|l|l|l}
            Legend: & \cellcolor{c00} 0-10\% & \cellcolor{c10} 10-20\% & \cellcolor{c20} 20-30\% & \cellcolor{c30} 30-40\% & \cellcolor{c40} 40-50\% & \color{white}\cellcolor{c50} 50-60\% & \color{white}\cellcolor{c60} 60-70\% & \color{white}\cellcolor{c70} 70-80\% & \color{white}\cellcolor{c80} 80-90\% & \color{white}\cellcolor{c90} 90-100\% \\
        \end{tabular}
        \begin{tabular}{l}
            (percentage of survey respondents mentioning a channel being used for an activity)
        \end{tabular}
    \end{center}
    \caption{Preliminary knowledge activities and the channels developers in our survey use for them.}
\label{table:survey:channels}
    \vspace{0pt}
\end{table}

\subsection{Communication Channels Developers Use to Support Their Development Activities}
\label{ssub:tse}
\begin{description}
\item \textit{This work is under review for a TSE journal - we described part of it in Section~\ref{sub:dev_challenges}}
\end{description}

In the study described in Section~\ref{sub:dev_challenges} we also inquired on the purpose of using each channel---i.e., which channels used for what activity. Table~\ref{table:survey:channels} shows an overview of which channels developers said they use to support different activities. This is a preliminary list of knowledge activities based on previous studies and a literature review. In this table, we have grouped the channels according to \emph{analog}, \emph{digital} and \emph{social+digital}. We can see from this table that there appears to be more reliance on communication channels that support social features.

\vspace{5mm}

In an effort to refine the preliminary knowledge activities (Table~\ref{table:survey:channels}), we designed the a follow-up study (described next), focusing on the knowledge activities developer do. These activities in fact foster the knowledge flow in the development process (as shown in Fig.~\ref{fig:knowledge_diagram}).

\subsection{A Typology of Knowledge Activities in Software Development} % (fold)
\label{sub:developer_activities}
\begin{description}
\item \textit{This work has been partially completed}
\end{description}

\begin{figure}[h!]
    \begin{center}
        \includegraphics[width=\textwidth]{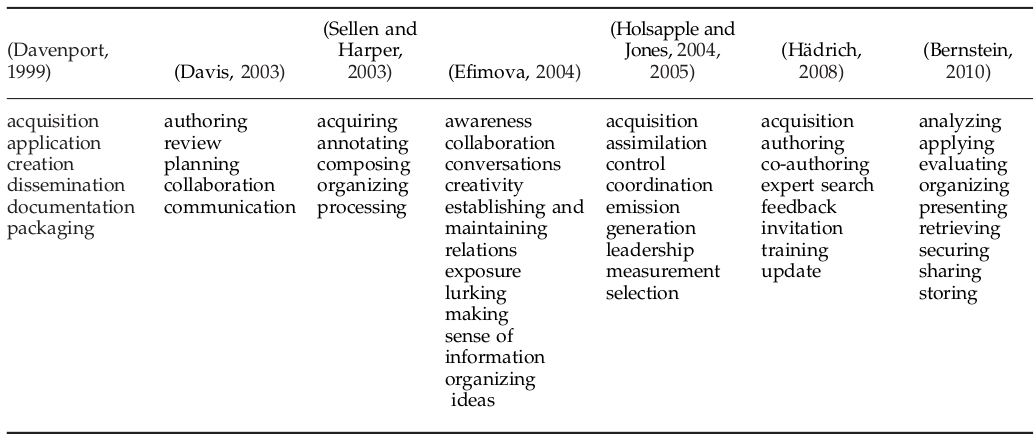}
        \caption{Existing taxonomies for knowledge actions, as depicted by Reinhardt \textit{et al.}~\cite{reinhardt2011knowledge}.}
        \label{fig:activities}
    \end{center}
\end{figure}

One of our main goals in the proposed thesis is to create a typology of knowledge activities in software development. Reinhardt \textit{et al.}~\cite{reinhardt2011knowledge} have described a knowledge worker activity typology (which he called actions), that he based on previous work (see Figure~\ref{fig:activities}) and an empirical study (involving a task execution study and a survey). Developers are a knowledge worker type (some even say developers are the knowledge worker prototype), however, Reinhardt \textit{el al.}'s typology is insufficient in capturing the activities done in software development and so does other knowledge work activity taxonomies. As a result, we aim to form and validate a typology of knowledge activities for developers.

Based on an iterative process of \textit{data analysis}---\textit{discussion}---\textit{literature review}, we came up with the following initial activity typology. Note that the preliminary activities from the previous study are all represented below (e.g., \textit{``Watch Activities''} is now \textit{``Monitoring''}):

\begin{enumerate}[noitemsep]
    \item \textbf{Acquisition}: gathering of information with the goal of developing skills or project or obtaining an asset
    \item \textbf{Authoring}: the creation code and related artifacts (externalized knowledge)
    \item \textbf{Co-Authoring}: the collaborative creation of textual and medial content (externalized knowledge)
    \begin{enumerate}[noitemsep]
        \item Communicating with others (including users)
        \item Coordinate tasks with others
    \end{enumerate}
    \item \textbf{Dissemination}: can be either of content (e.g., code or documentation) or of developer's activity (e.g., status reports)
    \begin{enumerate}[noitemsep]
        \item Dissemination of content
        \item Displaying activities
    \end{enumerate}
    \item \textbf{Feedback}: includes both gaining and providing feedback
    \item \textbf{Information Organization}
    \item \textbf{Learning}: acquiring new knowledge, skills, or understanding during the execution of work or based on formalized learning material
    \begin{enumerate}[noitemsep]
        \item Ask and answer questions
        \item Using examples
    \end{enumerate}
    \item \textbf{Monitoring}: includes the following
    \begin{enumerate}[noitemsep]
        \item Staying up to date: stay up to date about new technologies, practices, trends and tools for software development
        \item Keep track of one's development activities
        \item Monitor one's collaborators' activities
        \item Serendipitous discovery
    \end{enumerate}
    \item \textbf{Networking}: creating and maintaining a (professional?) social network
    \item \textbf{Searching}: looking up information, services, or experts
\end{enumerate}

\subsubsection{Methodology}
As a first phase, we conducted in-depth interviews with developers asking \textbf{what} activities they do, \textbf{why} and \textbf{how}, what \textbf{channels and tools} they use for each activity, and what \textbf{challenges} they face. So far we interviewed 7 participants (with a 28\% response rate) and transcribed 3 of these interviews. The interviews were 70-90 minutes long, and were conducted over VOIP (e.g., Skype, Google Hangouts). Note that this builds on the findings of previous studies (described in Sections ~\ref{ssub:fose2014} and \ref{ssub:tse}), which helped form the initial set of knowledge activities. Thus, in this study we use interviews as their strength lies in the ability to gain a deeper understanding of each activity.

\subsubsection{How Does This Fit in The Thesis}
The knowledge activity typology for software development would allow us to categorize and map the knowledge flow, the channel use, and the developer roles in software development. The interviews conducted so far support the initial activity typology we built, in fact, none of the interviews revealed any new activities so far.

\subsubsection{What is Next?}
First we have to finish the first phase, conduct additional interviews with developers, and analyze these interviews (this work in under progress now).

The next phase will be to conduct observations of developer teams and recording the activities done in \textit{the wild} (the subject of the observations is to be determined). This research method will complement our previous phase, as it would allow us to overcome the limitations associated with interviews (a possibility of bias and the risk of capturing only what we asked about).

% subsection developer_activities (end)
\newpage
\section{Towards a Participatory Process in SE Research}
% \begin{description}
% \item \textit{This work has been published as a keynote at EASE 2015 conference}
% \end{description}

Selecting suitable research methods for empirical software engineering research is ambitious due to the various benefits and challenges each method entails. One must regard the different theoretical stances relating to the nature of the research questions being asked, as well as practical considerations in the application of methods and data collection~\cite{Easterbrook}.

In this study~\cite{Storey2015ease}, we described some of the characteristics of this emergent development culture and the tools that are used to support it. We also discussed the different tradeoffs and challenges faced in selecting and applying research methods for studying development in online communities of practice. 
Finally, we briefly explore how social tools could foster a more participatory culture in software engineering research and how that may help to accelerate our impact on software engineering practice. 

In our studies of this participatory development culture, our goal has not been to investigate or experiment with new tools or ideas, but rather to understand the impact of the social tools developers already adopt on their development practices and on the community. We discuss two approaches, \textbf{mining software repositories} and \textbf{burrowing methods from social sciences}. We then describe how and why we \textbf{blend both} in our research.

Many software engineering researchers have already established excellent \textbf{social media literacy skills}.
We have used our social graph to \textbf{gain research data and insights}, and to \textbf{form alliances} with developers and collaborations with other researchers. We also find it a useful avenue for \textbf{disseminating our research} and learning about related research in our field. It also becomes a useful \textbf{backchannel} for discussing research as it is being presented at conferences.

However, there are many challenges and risks from using social tools in our research community. We mentioned the need for researchers to develop improved \textbf{media literacy skills}, otherwise they may miss out on research developments. Another risk is that easier access to participants through public channels (such as GitHub) may lead to us \textbf{inadvertently spamming} our participants. Another issue we need to be aware of is to ensure that we do not somehow \textbf{cause harm to our participants}.

\section{The Emergence of The Participatory Process in SE Education}
\begin{description}
\item \textit{This work has been published at CSCW 2015 conference, and a follow-up\\study is under review for ICSE 2016, SEET track}
\end{description}
In this study, we examined how GitHub is emerging as a collaborative platform for education. We aim to understand how environments such as GitHub---environments that provide social and collaborative features in conjunction with distributed version control---may improve (or possibly hinder) the educational experience for students and teachers. We conducted a qualitative study focusing on how GitHub is being used in education, and the motivations, benefits and challenges it brings.

\subsubsection{Methodology}
To discover if and how GitHub is being used in education, our study used exploratory research methods \cite{easterbrook2008selecting,seaman1999qualitative} and consisted of three phases of data collection (online methods \cite{wakeford2008fieldnotes}, interviews, and a validation survey). We were guided by the following research questions:

\begin{enumerate}[noitemsep]
    \item[\textbf{SRQ1}] How does GitHub support learning and teaching?
    \item[\textbf{SRQ2}] What are the motivations and benefits of using GitHub for education?
    \item[\textbf{SRQ3}] What challenges are related to the use of GitHub for education?
\end{enumerate}

\subsubsection{The GitHub Participatory Workflow}
We found that GitHub can be compared to traditional learning management systems (LMS), however, GitHub allows going beyond the traditional LMS and provides many benefits when used to support teaching and learning. Benefits include: providing transparency of activity, encouraging participation, allowing reuse and sharing of knowledge (and course material), providing industry relevance, and provides a shared space for work (thus supporting a community of practice). We also noted challenges educators faced when using GitHub in the classroom: lack of a shared knowledge base, barriers to entry, and external restrictions (e.g., copyright).

\subsubsection{The Student's Perspective}
From our previous phase, we found that educators leverage GitHub's collaboration and transparency features to create, reuse and remix course materials, and to encourage student contributions and monitor student activity on assignments and projects. However, our previous research did not consider the student perspective. In this phase, we conducted a case study where GitHub was used as a learning platform for two software engineering courses. We gathered student perspectives on how the use of GitHub in their courses might benefit them and to identify the challenges they may face.

The research questions addressed in this work include:
\begin{enumerate}[noitemsep]
    \item[\textbf{SRQ1:}] How do students benefit from using GitHub in their courses?
    \item[\textbf{SRQ2:}] What challenges do students face when GitHub is used by software engineering course instructors?
    \item[\textbf{SRQ3:}] What recommendations can we provide software engineering instructors who wish to use GitHub for teaching?
\end{enumerate}

The findings indicate that software engineering students do benefit from GitHub's transparent and open workflow. However, students were concerned that since GitHub is not inherently an educational tool, it lacks key features important for education and poses learning and privacy concerns. Our findings provide recommendations for designers on how tools such as GitHub can be used to improve software engineering education, and also point to recommendations for instructors on how to use it more effectively in their courses.

\subsubsection{How Does This Fit in The Thesis}
GitHub is on the brink of growing from a platform for software projects, and into a mainstream collaboration platform for other domains as well. An unexpected area where GitHub’s collaborative workflow holds the potential to bring groundbreaking changes is education and learning by promoting a participatory culture. In fact, as we seen educators have already begun to use GitHub to support teaching and learning. Moreover, Slack is also being adopted in supporting education as well\footnote{\url{http://www.sqrlab.ca/blog/2015/11/29/using-slack-in-the-classroom/}}.

Software engineering education is an important part of the knowledge theory process in software development, and there is a bi-directional relation between the two. From one perspective, developers are on a life long journey of learning and knowledge acquisition. While on the other hand, the practice of software development (e.g., practitioners) feed back into software engineering educational system. In fact, communities of practice foster both sides of the eco-system (e.g., communities on Stack Overflow).

% section collaboration (end)

% section activity_theory (end)
%!TEX root = candidacy.tex

\section{Discussion} % (fold)
\label{sec:discussion}
We described an emerging theory of knowledge in software engineering, and the way we understand it through our studies of the participatory process in software development. Figure~\ref{fig:knowledge_diagram} represents the knowledge theory as we understand it now, and as this is a work in progress, we continue developing and shaping this theory. However, few questions come to mind.

\subsubsection{Why Do We Need Another Theory}
We have Wasko's~\cite{Wasko2000} theories of knowledge, Reinhardt \textit{et al.}'s~\cite{reinhardt2011knowledge,reinhardt2012awareness} knowledge worker theory, and Naur's theory~\cite{naur1985programming}---why do we need another theory?
These theories do not provide the sufficient tools and terminology to study the participatory process in software development. Moreover, we don't intend to replace these theories, but expand and build on their work. Bringing these theories and findings from the different studies is one of the contributions of the proposed thesis.

\subsubsection{Is This A Theory or A Meta-Theory}
A good question. For example, can this theory be used to analyze the Agile process as well (meta-theory) or does it only describe the participatory process (theory)? Perhaps yes. This knowledge theory has emerged from our study of the participatory process in software development, and the components we describe (Fig.~\ref{fig:knowledge_diagram}) are needed in order to study the participatory culture. However, this work may be generalizable in the future, by separating between the components and the specific instance we study (i.e., the participatory process).

\subsubsection{Limitations}
We discuss limitations and threats to validity separately in each study (not described here), however, there is one concern that should be mentioned here.
Three of the studies described earlier (Sections~\ref{ssub:fose2014}, \ref{ssub:tse}, and \ref{sub:developer_activities}) have used the same community to recruit participants, i.e., in all three studies we recruited participants that are active on GitHub (though \textbf{different} and \textbf{independent} participants each time). We used this community by choice, as GitHub is widely used by software teams. However, this may limit the generalizability of our findings, as perhaps developers who are active on GitHub may engage more in the participatory process. This concern may be addressed in the future, by expanding or changing the participation community.

% section discussion (end)
%!TEX root = candidacy.tex

\section{Timeline} % (fold)
The work described in the previous chapters has been conducted in the first half of my thesis (2 years). The work described here was not done solely by myself, it has been done in collaboration with other researchers, and my contribution varies between the studies. Below is a tentative timeline for the completion of the activities leading to the final dissertation.

\begin{itemize}
    \item Continue study on developer activities (incl. interviews) and form a knowledge activity typology - finish by August 2016.
    \item Build a framework of the Participatory Process in software development - by January 2017.
    \item Demonstrate the use of the framework on a developer community (e.g., SendWithUs or Neo4J) - by August 2018.
    \item Write up dissertation - by September 2018.
\end{itemize}

% section timeline (end)

\end{onehalfspacing}

\newpage
\bibliographystyle{plain} 
\bibliography{candidacy}

\end{document}